\documentclass[a4paper,12pt]{article}
\usepackage{amssymb,amsmath}
\usepackage{fancyhdr,verbatim}

\usepackage{a4}
\usepackage{parskip}
\usepackage{appendix}
\usepackage{color,soul,rotating}
\usepackage{amsthm,amsxtra,amscd,relsize,multirow}
\usepackage[all,2cell]{xy}\UseAllTwocells

\newcommand{\sect}[1]{\setcounter{equation}{0}\section{#1}}

\def\N{{\mathcal N}}

\def\V{{\mathrm V}}

\def\sinh{\mathrm{sinh}}
\def\cosh{\mathrm{cosh}}

\def\axs{AdS_5\times S^5}
\newcommand{\eq}[1]{\begin{equation} #1 \end{equation}}
\newcommand{\al}[1]{\begin{align} #1 \end{align}}

\begin{document}
\begin{titlepage}
\markright{\bf TUW--12--11}
\title{On semiclassical four-point correlators in $AdS_5\times S^5$}

\author{D.~Arnaudov${}^{\star}$ and R.~C.~Rashkov${}^{\dagger,\star}$\thanks{e-mail:
rash@hep.itp.tuwien.ac.at}
\ \\ \ \\
${}^\star$  Department of Physics, Sofia
University,\\
5 J. Bourchier Blvd, 1164 Sofia, Bulgaria
\ \\ \ \\
${}^\dagger$ Institute for Theoretical Physics, \\ Vienna
University of Technology,\\
Wiedner Hauptstr. 8-10, 1040 Vienna, Austria
}
\date{}
\end{titlepage}

\maketitle
\thispagestyle{fancy}

\begin{abstract}
Following the recent advances in the holographic calculation of $n$-point correlation functions with two ``heavy'' (with large quantum numbers) states at strong coupling, we extend these findings by computing specific four-point correlators of four heavy BMN operators in $\N=4$ SYM.
\end{abstract}

\sect{Introduction}

For the last forty years there have been attempts to realize a correspondence between gauge theories with large number of colors and string theory. The most convincing duality of this kind was formulated relatively recently, when Maldacena conjectured the AdS/CFT correspondence \cite{Maldacena}. Many compelling results originating from the duality between type-IIB superstring theory on $AdS_5\times S^5$ and ${\cal N}=4$ supersymmetric Yang-Mills theory~\cite{Maldacena,GKP,Witten} established this field as a major research area.

One of the results of this duality is the relation between planar correlation functions of single-trace conformal primary operators in the boundary gauge theory and correlators of corresponding closed-string vertex operators on a worldsheet with the topology of $S^2$. Let us start with the study of a correlation function with two heavy vertex operators (with large quantum numbers, proportional to the string tension) and a number of ``light'' vertex operators (with quantum numbers and dimensions close to one). Then the large $\sqrt{\lambda}$ behavior of correlators of such operators is determined by a semiclassical string trajectory governed by the heavy operator insertions, while light vertex operators play the role of sources. To be more precise, first we find the classical string solution that determines the leading contribution to the two-point function of the heavy operators. Then we calculate the complete correlator by evaluating the light vertex operators on this trajectory.

The above semiclassical considerations were utilized for the calculation of two-point functions in \cite{Polyakov:2002}--\cite{Buchbinder:2010vw}. An extension of this method to certain three-point correlators was discussed in \cite{Janik:2010gc,Buchbinder:2010vw}, and elaborated in \cite{Zarembo:2010,Costa:2010}, where the heavy operators corresponded to a semiclassical string state with large R-charge and the light operator represented a BPS state -- massless (supergravity) scalar or dilaton mode. A more refined approach based on vertex operator insertions was put forward in \cite{Roiban:2010}. Further studies can be found in~\cite{Hernandez:2010}--\cite{Caputa:2012}. At present the main efforts of the researchers worldwide are concentrated on the calculation of three-point functions of three heavy operators \cite{Klose:2011}--\cite{Kazama:2012}.

Parallel to these advances, an extension to four-point correlators was initiated in \cite{BT:2010,Arnaudov:2011} and various correlation functions were computed. Furthermore, comparison with results from the gauge theory side was provided. Motivated by these studies, in the present paper we examine the leading in $\sqrt{\lambda}$ contributions to particular four-point functions of four heavy BMN operators from the point of view of string theory in $\axs$. The correlators can be represented by an exchange diagram involving a chiral primary operator (CPO).

The paper is organized in the following way. Sec. 2 is devoted to a brief overview of the method for evaluating $n$-point correlation functions with vertex operators. In Sec. 3 we proceed with the calculation of a leading contribution to four-point correlators of four BMN operators. We briefly discuss the results in the Conclusion.

\sect{Correlation functions with two heavy operators}

First we will review the case of two-point correlators. In the leading semiclassical approximation they are determined by the corresponding classical string solution \cite{Tseytlin:2003}--\cite{Janik:2010gc}. If $V_{H1}(\xi_1)$ and $V_{H2}(\xi_2)$ are the two heavy vertex operators inserted at points $\xi_1$ and $\xi_2$ on the worldsheet, the two-point function for large string tension ($\sqrt{\lambda}\gg1$) is given by the stationary point of the action
\eq{
\langle V_{H1}(\xi_1)V_{H2}(\xi_2)\rangle\sim e^{-I},
}
where $I$ represents the action of the $\axs$ string sigma model in the embedding coordinates
\al{
&I=\frac{\sqrt{\lambda}}{4\pi}\int d^2\xi\ \Big(\partial Y_M\bar{\partial}Y^M+\partial X_k\bar{\partial}X_k+{\rm fermions}\Big)\,,\\
&Y_MY^M=-Y_0^2-Y_5^2+Y_1^2+Y_2^2+Y_3^2+Y_4^2=-1\,,\quad\ X_kX_k=X_1^2+\ldots+X_6^2=1\,.\nonumber
}
We will impose conformal gauge and assume that the worldsheet is endowed with Euclidean signature. The 2D derivatives have the following form $\partial=\partial_1+i\partial_2,\,\bar{\partial}=\partial_1-i\partial_2$. The relation between the embedding coordinates, and the global and Poincar\'e coordinates of $AdS_5$ that we will use below is
\al{
&Y_5+iY_0=\cosh\,\rho\ e^{it},\quad
Y_1+iY_2=\sinh\,\rho\,\cos\theta\,e^{i\phi_1},\quad
Y_3+iY_4=\sinh\,\rho\,\sin\theta\,e^{i\phi_2},\nonumber\\
&Y_m=\frac{x_m}{z}\,,\qquad
Y_4=\frac{1}{2z}(-1+z^2+x^mx_m)\,,\qquad
Y_5=\frac{1}{2z}(1+z^2+x^m x_m)\,,
\label{poincare}
}
where $x^mx_m=-x_0^2+x_ix_i\ (m=0,1,2,3;\ i=1,2,3)$. However, our approach necessitates the use of the Euclidean continuation of $AdS_5$
\eq{
t_e=it\,,\qquad Y_{0e}=iY_0\,,\qquad x_{0e}=ix_0\,,
}
so that $Y_MY^M=-Y_5^2+Y_{0e}^2+Y_iY_i+Y_4^2=-1$.

The stationary-point solution solves the string equations of motion with singular sources provided by $V_{H1}(\xi_1)$ and $V_{H2}(\xi_2)$. Making use of the conformal symmetry of the theory, it is possible to map the $\xi$-plane to the Euclidean cylinder parameterized by $(\tau_e,\sigma)$
\eq{
e^{\tau_e+i\sigma}=\frac{\xi-\xi_2}{\xi-\xi_1}\,.
\label{confmap}
}
This conformal map transforms the singular solution on the $\xi$-plane to a smooth string solution \cite{Tseytlin:2003,Buchbinder:2010,Buchbinder:2010vw}. The new solution has the same quantum numbers as the vertex-operator states, guaranteeing that no information is lost.

The above considerations can be reformulated for a physical integrated vertex operator labeled by a point ${\rm x}$ on the boundary of $AdS_5$ \cite{Polyakov:2002,Tseytlin:2003}
\eq{
{\rm V}_H({\rm x})=\int d^2\xi\ V_H(\xi;{\rm x})\,,\qquad V_H(\xi;{\rm x})\equiv V_H(z(\xi),x(\xi)-{\rm x},X_k(\xi))\,.
}
The two-point function $\langle\V_{H1}({\rm x}_1)\V_{H2}({\rm x}_2)\rangle$ is determined in a similar fashion by the classical action evaluated on the stationary point solution. After applying the conformal map to the cylinder~\eqref{confmap} we obtain a smooth solution that is actually the corresponding solitonic string solution in terms of Poincar\'e coordinates which satisfies the boundary conditions~\cite{Buchbinder:2010vw}
\eq{
\tau_e\rightarrow-\infty\ \ \Longrightarrow\ \ z\rightarrow0\,,\ \ x\rightarrow{\rm x}_1\,,\qquad\qquad\tau_e\rightarrow+\infty\ \ \Longrightarrow\ \ z\rightarrow0\,,\ \ x\rightarrow{\rm x}_2\,.
\label{boundcond}
}

As was shown in \cite{Roiban:2010}, the semiclassical three-point correlators with two heavy and one light operators are of the form
\al{
G_3({\rm x}_1,{\rm x}_2,{\rm x}_3)&=\langle\V_{H1}({\rm x}_1)\V_{H2}({\rm x}_2)\V_L({\rm x}_3)\rangle\\
&=\int{\cal D}\mathbb{X}^\mathbb{M}\ e^{-I}\!\int d^2\xi_1d^2\xi_2d^2\xi_3\ V_{H1}(\xi_1;{\rm x}_1)V_{H2}(\xi_2;{\rm x}_2)V_L(\xi_3;{\rm x}_3)\,,\nonumber
}
where $\int{\cal D}\mathbb{X}^\mathbb{M}$ is the integral over $(Y_M,X_k)$. In the stationary point equations the contribution of the light operator can be neglected, so that the solution coincides with the one in the case of the two-point function of the two heavy operators. Thus we get \cite{Roiban:2010}
\eq{
\frac{G_3({\rm x}_1,{\rm x}_2,{\rm x}_3)}{G_2({\rm x}_1,{\rm x}_2)}=\int d^2\xi\ V_L(z(\xi),x(\xi)-{\rm x}_3,X_k(\xi))\,,
\label{strconstxi}
}
where $(z(\xi),x(\xi),X_k(\xi))$ stands for the corresponding string solution with the same quantum numbers as the heavy vertex operators, transformed to the $\xi$-plane by \eqref{confmap}. Taking account of the fact that $\int d^2\sigma=\int^\infty_{-\infty}d\tau_e\int^{2\pi}_0d\sigma$, we can also represent \eqref{strconstxi} in terms of the cylinder
\eq{
\frac{G_3}{G_2}=\int d^2\sigma\ V_L(z(\tau_e,\sigma),x(\tau_e,\sigma)-{\rm x}_3,X_k(\tau_e,\sigma))\,.
\label{strconstst}
}

The global conformal $SO(2,4)$ symmetry fixes the two- and three-point correlation functions (assuming that $\V_{H2}=\V^*_{H1}$)
\al{\label{2point}
G_2({\rm x}_1,{\rm x}_2)&=\frac{C_{12}\ \delta_{\Delta_1\!,\Delta_2}}{{\rm x}_{12}^{\Delta_1+\Delta_2}}\,,\qquad{\rm x}_{ij}\equiv|{\rm x}_i-{\rm x}_j|\,,\\
G_3({\rm x}_1,{\rm x}_2,{\rm x}_3)&=\frac{C_{123}}{{\rm x}_{12}^{\Delta_1+\Delta_2-\Delta_3}{\rm x}_{13}^{\Delta_1+\Delta_3-\Delta_2}
{\rm x}_{23}^{\Delta_2+\Delta_3-\Delta_1}}\,,
\label{3point}
}
where $\Delta_i$ are the scaling dimensions of the operators. If we choose in a convenient way ${\rm x}_i$, the dependence on ${\rm x}_{ij}$ in \eqref{strconstst} can be removed. Then \eqref{strconstst} allows to compute the structure constants $C_{123}$~\cite{Roiban:2010}. Presuming that $\Delta_1=\Delta_2$, we find after setting $C_{12}=1$ in \eqref{2point} that
\eq{
\frac{G_3({\rm x}_1,{\rm x}_2,{\rm x}_3=0)}{G_2({\rm x}_1,{\rm x}_2)}=C_{123}\left(\frac{{\rm x}_{12}}{|{\rm x}_1|\,|{\rm x}_2|}\right)^{\Delta_3}\!.
\label{norm3point}
}

The leading contribution to the four-point functions of two heavy and two light operators is provided by
\al{
&G_4({\rm x}_1,{\rm x}_2,{\rm x}_3,{\rm x}_4)=\langle\V_{H1}({\rm x}_1)\V_{H2}({\rm x}_2)\V_{L1}({\rm x}_3)\V_{L2}({\rm x}_4)\rangle\\
&=\int{\cal D}\mathbb{X}^\mathbb{M}\ e^{-I}\!\int d^2\xi_1d^2\xi_2d^2\xi_3d^2\xi_4\ V_{H1}(\xi_1;{\rm x}_1)V_{H2}(\xi_2;{\rm x}_2)
V_{L1}(\xi_3;{\rm x}_3)V_{L2}(\xi_4;{\rm x}_4)\,.\nonumber
}
The semiclassical trajectory being the same, $G_4$ is determined by the product of the light operators on the solution
\eq{
\frac{G_4({\rm x}_1,{\rm x}_2,{\rm x}_3,{\rm x}_4)}{G_2({\rm x}_1,{\rm x}_2)}=\!\int\!d^2\xi_3\,V_{L1}(z(\xi_3),x(\xi_3)-{\rm x}_3,X_k(\xi_3))\!\!\int\!d^2\xi_4\,V_{L2}(z(\xi_4),x(\xi_4)-{\rm x}_4,X_k(\xi_4))\,.
}
Transforming to the $(\tau_e,\sigma)$-coordinates we get
\eq{
\frac{G_4}{G_2}=\!\int\!d^2\sigma d^2\sigma'\ V_{L1}(z(\tau_e,\sigma),x(\tau_e,\sigma)-{\rm x}_3,X_k(\tau_e,\sigma))\, V_{L2}(z(\tau'_e,\sigma'),x(\tau'_e,\sigma')-{\rm x}_4,X_k(\tau'_e,\sigma'))\,.
\label{4point}
}

\sect{Four-point functions of heavy BMN states}

\begin{figure}[htb]
\begin{center}
\includegraphics[width=5cm]{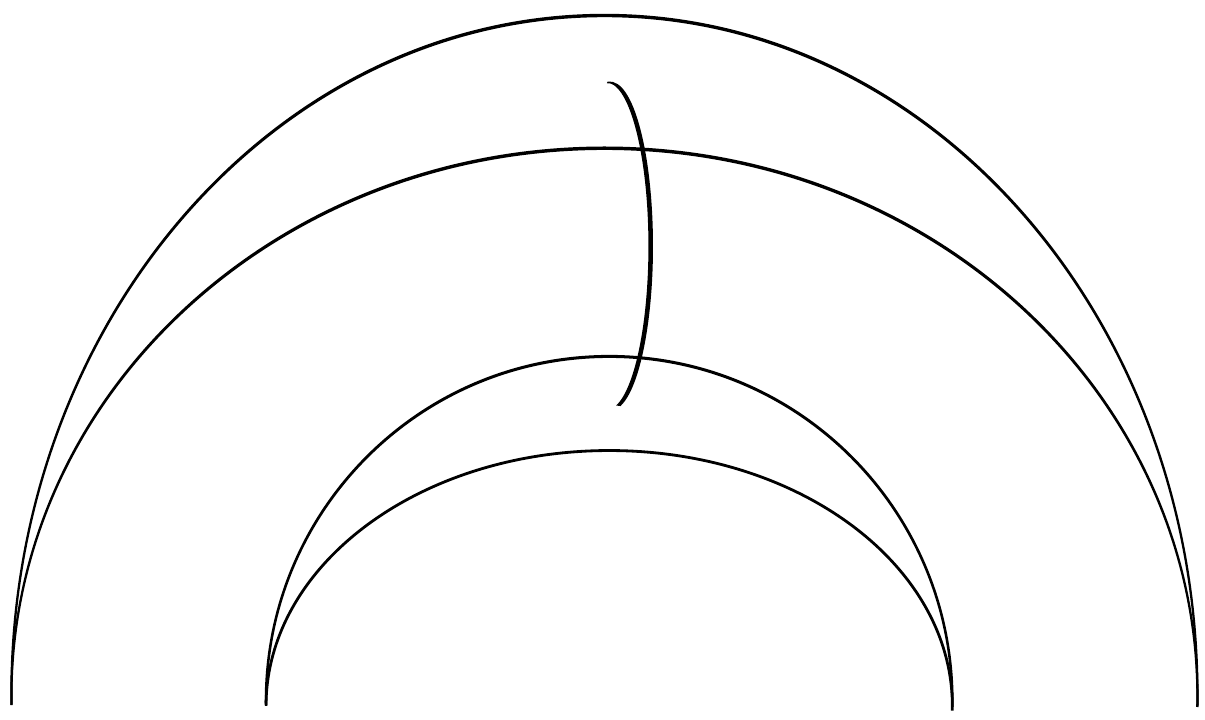}
\end{center}
\caption{\label{4pheavy} Witten diagram for the four-point function of four heavy BMN states.}
\end{figure}

In this Section we will consider particular constrained contributions to four-point correlators of four heavy BMN operators corresponding to the string solution in \cite{Buchbinder:2010vw} for generic positions of the heavy operator insertions. The correlation function can be thought of as two two-point correlators connected with a light bulk-to-bulk propagator (see fig. \ref{4pheavy}). To obtain the full correlation functions we need to sum over all intermediate supergravity states. We will follow the spirit of the calculations presented in \cite{BT:2010}.

The large-spin gauge theory operators in the correlator at hand are dual to the point-like in AdS strings with one angular momentum in $S^5$ \cite{Buchbinder:2010vw}. The positions of the heavy operators on the boundary are chosen in the form ${\rm x}_1=-{\rm x}_4$ and ${\rm x}_2=-{\rm x}_3$. The corresponding Euclidean stationary point solutions for the two semiclassical propagators in fig. \ref{4pheavy} are \cite{BT:2010}\footnote{We choose the points ${\rm x}_i,\,i=1,\dots,4$, to lie on the $x_{0e}$-axis.}
\begin{subequations}
\label{x12sol}
\al{
z&=\frac{{\rm x}_{14}}{2\cosh(\kappa\tau_e)}\,,\quad x_{0e}=\frac{{\rm x}_{14}}{2}\tanh(\kappa\tau_e)\,,\quad x_i=0\,,\quad
\varphi_1=-i\kappa\tau_e\,,\quad\kappa=\frac{J}{\sqrt{\lambda}}\,,\\
z'&=\frac{{\rm x}_{23}}{2\cosh(\kappa'\tau'_e)}\,,\quad x'_{0e}=\frac{{\rm x}_{23}}{2}\tanh(\kappa'\tau'_e)\,,\quad x'_i=0\,,\quad
\varphi'_1=-i\kappa'\tau'_e\,,\quad\kappa'=\frac{J'}{\sqrt{\lambda}}\,,
}
\end{subequations}
and we assume without loss of generality that ${\rm x}_{14}\geq{\rm x}_{23}$. We denote the respective heavy vertex operators as $\V_J$ and $\V_{J'}$, with $\V_{-J}\equiv\V^*_J$ and $\V_{-J'}\equiv\V^*_{J'}$. The two-point function of such operators can be calculated as \cite{Buchbinder:2010,Buchbinder:2010vw}
\eq{
\langle\V_{-J}({\rm x}_1)\V_J({\rm x}_4)\rangle=\frac{1}{{\rm x}_{14}^{2\Delta(J)}}\,,\quad
\langle\V_{-J'}({\rm x}_2)\V_{J'}({\rm x}_3)\rangle=\frac{1}{{\rm x}_{23}^{2\Delta(J')}}\,,\quad
\Delta(J)=J\,,\quad\Delta(J')=J'.
}
In order to obtain a contribution to the four-point correlator we need to ``join'' two two-point functions with a light state, which we choose to be the chiral primary operator. The massless string state corresponding to the CPO originates from the trace of the graviton in the $S^5$ directions \cite{Kim:1985,Lee}. Building on results in \cite{Zarembo:2010,Berenstein:1998}, we conjecture that the bosonic part of the respective supergravity ``vertex operator'' has the following form\footnote{We ignore derivative terms that will not influence the calculation below since we work with ${\rm x}_1=-{\rm x}_4$ and ${\rm x}_2=-{\rm x}_3$.}
\al{
&{\bf V}_L(z,x;z',x')={\bf V}^{(\rm CPO)}_{\Delta}(z,x;z',x')=\hat{c}_{\Delta}^2G_{\Delta}(z,x;z',x')\,{\rm X}^j{\rm X'}^{-j}{\cal L}\,{\cal L}'\,,\qquad
\Delta=j\,,\\
&G_{\Delta}(z,z;z',x')=\frac{2^{-\Delta}(\Delta-1)\zeta^{\Delta}}{2\pi^2}\ {}_2F_1\!\left(\frac{\Delta}{2},\frac{\Delta+1}{2};\Delta-1;\zeta^2\right),\quad
\zeta=\frac{2zz'}{z^2+z'^2+(x-x')^2}\,,\nonumber\\
&{\rm X}\equiv X_1+iX_2=\sin\gamma\cos\psi\,e^{i\varphi_1},\qquad{\rm X'}\equiv X'_1+iX'_2=\sin\gamma'\cos\psi'\,e^{i\varphi'_1},\nonumber\\
&{\cal L}=\frac{\partial x_m\bar{\partial}x^m-\partial z\bar{\partial}z}{z^2}-\partial X_k\bar{\partial}X_k\,,\qquad
{\cal L}'=\frac{\partial'x'_m\bar{\partial}'x'^m-\partial'z'\bar{\partial}'z'}{z'^2}-\partial'X'_k\bar{\partial}'X'_k\,,
\nonumber
}
where the normalization constant $\hat{c}_\Delta$ of the CPO is given by \cite{Zarembo:2010,Berenstein:1998}
\eq{
\hat{c}_\Delta=\hat{c}_j=\frac{\sqrt{\lambda}}{8\pi N}(j+1)\sqrt{j}\,.
}
The light gauge theory operator is ${\rm Tr}Z^j$ with conformal dimension $\Delta=j\geq2$. The four-point correlator assumes the form
\eq{
\frac{G_4({\rm x}_1,{\rm x}_2,{\rm x}_3,{\rm x}_4)}{G_2({\rm x}_1,{\rm x}_4)G'_2({\rm x}_2,{\rm x}_3)}=\int d^2\xi d^2\xi'\,
{\bf V}^{(\rm CPO)}_{\Delta}(z,x;z',x')\,.
}
Evaluated on \eqref{x12sol}, it can be presented as
\eq{
\frac{G_4}{G_2G'_2}=16\pi^2\hat{c}_j^2\kappa^2\kappa'^2\int_{-\infty}^{\infty}d\tau_e\int_{-\infty}^{\infty}d\tau'_e\,
\frac{G_{\Delta}e^{j(\kappa\tau_e-\kappa'\tau'_e)}}{\cosh^2(\kappa\tau_e)\,\cosh^2(\kappa'\tau'_e)}\,.
\label{intheavy}
}
It can be shown that the bulk-to-bulk propagator for the CPO can be expressed in terms of elementary functions in the following way
\eq{
G_\Delta=\frac{\zeta^j[(1-\zeta^2)^{-1/2}+j-2]}{8\pi^2(1-\zeta^2)(1+\sqrt{1-\zeta^2})^{j-2}}\,.
}
Let us change the integration variables from $\tau_e$ and $\tau'_e$ to $y\equiv\tanh(\kappa\tau_e)$ and $y'\equiv\tanh(\kappa'\tau'_e)$. Then \eqref{intheavy} becomes
\eq{
\frac{G_4}{G_2G'_2}=2\hat{c}_j^2\kappa\kappa'\eta^j\int_{-1}^1dy\int_{-1}^1dy'\,
\frac{(1+y)^j(1-y')^j}{(1-\eta yy')^j}\frac{(1-\zeta^2)^{-1/2}+j-2}{(1-\zeta^2)(1+\sqrt{1-\zeta^2})^{j-2}}\,,
}
where we have introduced
$$
\eta=\frac{2{\rm x}_{14}{\rm x}_{23}}{{\rm x}^2_{14}+{\rm x}^2_{23}}\,.
$$
We finally obtain for the contribution to the four-point correlator
\al{
&\frac{G_4}{G_2G'_2}=2\hat{c}_j^2\kappa\kappa'\eta^j\int_{-1}^1dy\int_{-1}^1dy'\,[I_1+(j-2)I_2]\,,\\
&I_1=\frac{(1+y)^j(1-y')^j(1-\eta yy')}{P^{3/2}(1-\eta yy'+\sqrt{P})^{j-2}}\,,\qquad I_2=\frac{(1+y)^j(1-y')^j}{P(1-\eta yy'+\sqrt{P})^{j-2}}\,,
\nonumber
}
where $P=\eta^2(y^2+y'^2)-2\eta yy'+1-\eta^2$. If we choose for simplicity $j=2$, we will get
\eq{
G_4({\rm x}_1,{\rm x}_2,{\rm x}_3,{\rm x}_4)=\frac{C_{J,J'\!,\,j=2}}{{\rm x}_{14}^{2\Delta(J)}\,{\rm x}_{23}^{2\Delta(J')}}\,,
\label{4pbmn}
}
where
\al{\nonumber
C_{J,J'\!,2}&=2\hat{c}_2^2\kappa\kappa'\eta^2\int_{-1}^1dy\int_{-1}^1dy'\,\frac{(1+y)^2(1-y')^2(1-\eta yy')}{P^{3/2}}\\
&=\frac{9JJ'}{15\pi^2N^2}\!\left(\frac{(2\chi-1)(3\chi-1)}{2\chi}-\frac{(3\chi-2)\ln\chi}{\chi-1}\right)\!,\qquad\chi=\frac{({\rm x}_{14}-{\rm x}_{23})^2}{({\rm x}_{14}+{\rm x}_{23})^2}\,.
}
It can be easily seen that the right-hand expression in \eqref{4pbmn} behaves appropriately under conformal transformations. At first glance the ratio $\chi$ does not look conformally invariant, but if we account for the constraints ${\rm x}_1=-{\rm x}_4$ and ${\rm x}_2=-{\rm x}_3$ imposed on the positions of operators, we will see that $\chi$ and $\eta$ are not only invariant under translations, rotations and dilations, but also under inversion, which guarantees their conformal invariance. The existence of only one ratio (instead of two) follows from the constrained nature of operator positions.

\sect{Conclusion}

The duality between gauge and string theories passed through many crucial developments over the last years. The AdS/CFT correspondence has achieved impressive results about the anomalous dimensions of gauge theory operators, integrable structures, etc., and various properties of gauge theories at strong coupling have been established. One of the main challenges ahead is to find efficient methods for calculation of correlation functions.

Although the three-point correlator of three heavy operators is not yet fully understood~\cite{Klose:2011}--\cite{Kazama:2012}, we have discovered a lot about the behavior of correlation functions containing two heavy and one light states at strong coupling \cite{Zarembo:2010,Costa:2010}.\footnote{These considerations were initiated in \cite{Janik:2010gc}.} Using the trajectory for the correlator of two heavy operators \cite{Tseytlin:2003,Buchbinder:2010vw}, the authors of \cite{Roiban:2010} put forward an approach based on insertion of vertex operators. The same method applies also to higher $n$-point correlation functions with two heavy and $n-2$ light operators.

In the present paper we consider string theory on $\axs$ and compute specific leading contributions to four-point functions at strong coupling, applying the ideas of \cite{Roiban:2010} for calculation of correlators with vertex operators. We examine the method in the case of four heavy BMN states.

There has been also development in the computation of three- and four-point correlation functions with heavy operators via integrability techniques \cite{Escobedo:2010}--\cite{Grignani:2012b}. It would be very interesting to see if our results can be obtained in such a way from the gauge theory point of view.

\section*{Acknowledgments}
The authors would like to thank H. Dimov for valuable discussions and careful reading of the paper. This work was supported in part by the Austrian Research Funds FWF P22000 and I192, and NSFB DO 02-257.


\end{document}